\documentclass[aps,prd,reprint,10pt,preprintnumbers]{revtex4-1}
\usepackage{amsmath}
\usepackage{amsfonts}
\usepackage{graphicx}
\usepackage{subcaption}
\usepackage{caption}
\usepackage{comment}
\usepackage{toolbox}
\usepackage{xcolor}
\usepackage{braket}
\usepackage{hyperref}

\hypersetup{
	colorlinks = true,
	linkcolor = blue,
	citecolor = magenta
}

\usepackage{soul,xcolor}

\newcommand\cyansout{\bgroup\markoverwith{\textcolor{cyan}{\rule[0.5ex]{2pt}{0.4pt}}}\ULon}
\newcommand\blueout{\bgroup\markoverwith{\textcolor{blue}{\rule[0.5ex]{2pt}{0.4pt}}}\ULon}
\newcommand{\be}{\begin{equation}}
	\newcommand{\ee}{\end{equation}}
\newcommand{\bes}{\begin{equation*}}
	\newcommand{\ees}{\end{equation*}}
\newcommand{\bea}{\begin{eqnarray}}
	\newcommand{\eea}{\end{eqnarray}}
\newcommand{\beas}{\begin{eqnarray*}}
	\newcommand{\eeas}{\end{eqnarray*}}

\newcommand{\la}{\lambda}

\newcommand{\gmt}{(g-2)_\mu}

\newcommand{\lsim}{{\;\raise0.3ex\hbox{$<$\kern-0.75em\raise-1.1ex\hbox{$\sim$}}\;}}
\newcommand{\gsim}{{\;\raise0.3ex\hbox{$>$\kern-0.75em\raise-1.1ex\hbox{$\sim$}}\;}}

\begin{document}
	
	%\preprint{}
	
\title{Compatibility of muon $g-2$, $W$ mass anomaly in type-X 2HDM}
\author{Jongkuk Kim}
\affiliation{Korea Institute for Advanced Study, Seoul 02455, Korea}

\email{jongkuk.kim927@gmail.com}

\date{\today}
\begin{abstract}
Recently CDF II Collaboration reported that they measured W boson mass precisely. 
The measurement is deviated from the Standard Model (SM) Prediction at 7$\sigma$.
Also, the recent FNAL measurement of the muon magnetic moment shows a $4.2\sigma$ deviation.
To resolve the W boson as well as the muon $(g-2)$ anomalies, we explore the type-X two Higgs doublet model (2HDM). 
We analyze the $(m_A, \tan\beta)$ parameter space of the type-X 2HDM consistent with the muon 
$(g-2)$, lepton universality test and recent $W$ boson mass measurments from the CDF II collaboration. 
We find that the measurement of the $h\to AA$ branching ratio gives a strong lower mass bound on $m_H$.  
\end{abstract}

\preprint{KIAS-P220xx}
\date{\today}

%\keywords{}

\maketitle

%\subsection{}
%\subsubsection{}
\section{Introduction}
The CDF experiment presented a high precision measurement of mass of the $W$ boson:
$m_W^{CDF} = 80.4335\pm0.0094$ GeV, which is about $7\sigma$ away from the 
SM prediction \cite{CDF:2022hxs}. Such deveiation is in conflict expermental measurements
of $m_W$ by Tevatron \cite{CDF:2013dpa} and ATLAS \cite{ATLAS:2017rzl}. 
Undeniably, we should interpret the deviation with caution as it might originate from 
systematic errors.

On the other hand, this result stands with the growing number of experimental anomalies,
including the muon anomalous magnetic moment, $B$-anomalies etc. Hence it is worthwhile 
to reexamine the implication of the CDF result scenarios, which can explain other observed 
anomalies. The deviation in $W$ mass can originate from the quantum correction due to  
the presence of BSM particles which modifies the $S, T$ and $U$ parameters. 
Several analysises has been presented in this direction
\cite{ Zhu:2022tpr, Lu:2022bgw, Du:2022pbp, deBlas:2022hdk, Yang:2022gvz, Athron:2022qpo, Strumia:2022qkt, 
Cacciapaglia:2022xih, Liu:2022jdq, Sakurai:2022hwh, Athron:2022isz, Asadi:2022xiy, Song:2022xts, Bahl:2022xzi, 
Cheng:2022jyi, Lee:2022nqz, DiLuzio:2022xns, 
Gu:2022htv, Babu:2022pdn, Paul:2022dds, Biekotter:2022abc, 
Balkin:2022glu, DiLuzio:2022ziu, Cheung:2022zsb, Du:2022brr, Endo:2022kiw, 
Crivellin:2022fdf, Heo:2022dey, Ahn:2022xeq, Zheng:2022irz,
PerezFileviez:2022gmy,Ghoshal:2022vzo,Peli:2022ybi, Kawamura:2022uft, Kanemura:2022ahw, 
Nagao:2022oin, Mondal:2022xdy, Zhang:2022nnh,Carpenter:2022oyg, Popov:2022ldh, 
Arcadi:2022dmt, Chowdhury:2022moc,Borah:2022obi,Cirigliano:2022qdm,Zeng:2022lkk,Du:2022fqv, 
Ghorbani:2022vtv, Bhaskar:2022vgk,Baek:2022agi,Cao:2022mif,Borah:2022zim,Lee:2022gyf,
Almeida:2022lcs,Cheng:2022aau,Addazi:2022fbj,Heeck:2022fvl,Abouabid:2022lpg,Tan:2022bip,
Faraggi:2022emm,Batra:2022pej,Benbrik:2022dja,Cai:2022cti,
Dermisek:2022xal,Chen:2022ocr,Gupta:2022lrt}.

Here we are interested in the 2HDM of type-X,  which can explain the 
muon $(g-2)$ if the pseudoscalar ($A$) is much lighter than the other scalars. A considerable 
deviation in the $S$ and $T$ parameters is necessary to accommodate CDF results. We analyzed 
the parameter space which can simultaneously satisfy CDF measurement and $\gmt$. 
To explain $\gmt$, we require the pseudoscalar to be lighter than $m_h/2$ where $h$ is 
the SM Higgs bosons. And it opens up the $h\to AA$ decay model, which has been explored 
at LHC by CMS and ATLAS collaboration in various four-fermion final states, including 
$4\tau, 2\mu2\tau$ \cite{ATLAS:2015unc, CMS:2017dmg, ATLAS:2018coo,CMS:2018qvj, CMS:2019spf, CMS:2020ffa}. 
We showed that the LHC limit puts a strong lower limit on the 
mass spectrum of the type-X 2HDM. Since the scalars are leptophobic in type-X 2HDM, 
strong constraint comes from the measurement of lepton universality in charged and 
neutral current processes. We have considered both the leptonic $Z$ decays and 
leptonic/semi-leptonic $\tau$ decays to constrain the available parameter space. We 
showed that it is possible to explain the $\gmt$ anomaly after satisfying all the experimental and theoretical 
constraints, including the CDF measurement. We found that 
the masses of the heavy BSM scaalrs are strongly restricted from below, and can LHC and ILC can explore the parameters space.  

The paper is organized as follows:
In Sec.~\ref{sec:model}, we briefly summarize type-X 2HDM, muon $g-2$ contribution, lepton universality test and oblique parameters.
In Sec.~\ref{sec:result}, we describe all of the theoretical and experiment constraints and find preferred region.
We conclude in Sec.~\ref{sec:conclusion}.

\section{The Model Basics \label{sec:model} }
In this section we briefly describe the necessary elements of the type-X 2HDM. For a detailed discussions 
about multi Higgs doublet model please see~\cite{Gunion:1989we,Djouadi:2005gj,Branco:2011iw}.
\subsection{2HDM with type-X Yuakwa interaction}
The model contains two Higgs doublets $\Phi_1$ and $\Phi_2$ with $Y=1/2$. Presence of two Higgs doublets where 
both the doublets couple to the fermions leads to flavour changing neutral current (FCNC)
interaction at tree level. To avoid this, we impose a $\mathbb{Z}_2$ symmetry such that $\Phi_1\rightarrow-\Phi_1$ 
and $\Phi_2\rightarrow \Phi_2$. The scalar potential can be written as,
\begin{widetext}
\begin{eqnarray}
\nonumber V_{\mathrm{2HDM}} &=& -m_{11}^2\Phi_1^{\dagger}\Phi_1 - m_{22}^2\Phi_2^{\dagger}\Phi_2 -\Big[m_{12}^2\Phi_1^{\dagger}\Phi_2 + \mathrm{h.c.}\Big]
+\frac{1}{2}\lambda_1\left(\Phi_1^\dagger\Phi_1\right)^2+\frac{1}{2}\lambda_2\left(\Phi_2^\dagger\Phi_2\right)^2 \\
\nonumber && +\lambda_3\left(\Phi_1^\dagger\Phi_1\right)\left(\Phi_2^\dagger\Phi_2\right)+\lambda_4\left(\Phi_1^\dagger\Phi_2\right)\left(\Phi_2^\dagger\Phi_1\right)
+\Big\{ \frac{1}{2}\lambda_5\left(\Phi_1^\dagger\Phi_2\right)^2+  \rm{h.c.}\Big\}.
\label{eq:2hdm-pot}
\end{eqnarray}
\end{widetext}
After the electroweak symmetry breaking, the scalars $\Phi_1$ and $\Phi_2$ will acquire  vacuum expectation values (VEV) $v_1$ 
and $v_2$ respectively. We can parameterize the doublets in the following way,
$\Phi_j=(H_j^+,(v_j + h_j + i A_j)/\sqrt{2})^T$ and  obtain the physical states $A$, $h$, $H$, $H^{\pm}$ by appropriately 
rotating the gauge eigenstates:
\begin{align}\label{2hdm_scalar_basis}
 \begin{pmatrix} H  \\ h \end{pmatrix} &=  \begin{pmatrix}  c_{\alpha} && s_{\alpha} \\ -s_{\alpha} &&  c_{\alpha} \end{pmatrix}
  \begin{pmatrix} h_1  \\ h_2 \end{pmatrix} \nonumber\\
    A(H^{\pm})&=-s_\beta \;A_1(H_1^{\pm}) + c_\beta \;A_2(H_2^{\pm}), 
 \end{align}
where $s_\alpha = \sin\alpha$, $c_\beta = \cos\beta$ and $\tan\beta = v_2/ v_1$.
The CP-even state $h$ is identified with the SM-like Higgs with mass $m_h \approx 125$ GeV.

In this article we will consider the Type-X 2HDM where the RH leptons are odd under $\mathbb{Z}_2$ symmetry. 
The relevant Yukawa Lagrangian is given by,
\begin{equation}\label{eq:yukawa}
-\mathcal{L}_Y= Y^u\bar{ Q_L} \widetilde{ \Phi_2 } u_R + Y^d  \bar{ Q_L} \Phi_2 d_R+Y^e\bar{ l_L} \Phi_1 e_R + h.c.,
\end{equation}
where $\widetilde{ \Phi_2 }=i\sigma_2\Phi_2^*$. After symmetry breaking the Yukawa Lagrangian takes the form,
\begin{widetext}
\begin{eqnarray}
\nonumber \mathcal L_{\mathrm{Yukawa}}^{\mathrm{Physical}} &=&
-\sum_{f=u,d,\ell} \frac{m_f}{v}\left(\xi_h^f\overline{f}hf +
\xi_H^f\overline{f}Hf - i\xi_A^f\overline{f}\gamma_5Af \right) -\left\{ \frac{\sqrt{2}V_{ud}}
{v}\overline{u}\left(\xi_A^{u} m_{u} P_L+\xi_A^{d} m_{d} P_R\right)H^{+}d  +
\frac{\sqrt{2}m_l}{v}\xi_A^l\overline{v}_LH^{+}l_R + \mathrm{h.c.}\right\}.
\label{eq:L2hdm}
\end{eqnarray}
\end{widetext}
The Yukawa multiplicative factors, {\it i.e.} $\xi_{\phi}^f$ are given in Table~\ref{Tab:YukawaFactors}. 
\begin{table}[t]
%-------------------------------------------------------------------------------
\begin{center}
\begin{tabular}{|c||c|c|c|c|c|c|c|c|c|}
\hline
& $\xi_h^u$ & $\xi_h^d$ & $\xi_h^\ell$
& $\xi_H^u$ & $\xi_H^d$ & $\xi_H^\ell$
& $\xi_A^u$ & $\xi_A^d$ & $\xi_A^\ell$ \\ \hline
Type-X
& $c_\alpha/s_\beta$ & $c_\alpha/s_\beta$ & $-s_\alpha/c_\beta$
& $s_\alpha/s_\beta$ & $s_\alpha/s_\beta$ & $c_\alpha/c_\beta$
& $\cot\beta$ & $-\cot\beta$ & $\tan\beta$ \\
 \hline
\end{tabular}
\end{center}
 \caption{The multiplicative factors of Yukawa interactions in type X 2HDM}
\label{Tab:YukawaFactors}
\end{table}

The coupling modifier $\xi_h^\ell = -s_\alpha/c_\beta$ can be written as
$\sin(\beta-\alpha)-\tan\beta\cos(\beta-\alpha)$.

\subsection{Relevant Parameter Space for Muon $(g-2)$}
In the type-X 2HDM contribution to $\gmt$ comes at one loop diagram mediated by $H,A$ and $H^\pm$. 
The contribution from $H$ ix is positive, whereas the $A$ mediated process contributes negatively. The contribution to $\gmt$ coming
from $H^\pm$ diagram is insignificant. 
In addition, two-loop Barr-Zee diagram contributes to the $\gmt$ and dominated over the one-loop contribution.
At two-loop, $A(H)$ mediated process contributes positively(negatively). One can identify the parameter region 
of $(m_A, \tan\beta)$ plane consistent with the measurement of $a_\mu = (g-2)_\mu/2$ \cite{Muong-2:2021ojo}:
\begin{equation}
a_\mu(\mbox{Exp})-a_\mu(\mbox{SM})=(251\pm59) \times 10^{-11}.
\end{equation}
The pseudoscalar has to be sufficiently light, and a large enough enhancement on the leptonic 
coupling is necessary, which requires a large $\tan\beta$. So for our numerical analysis, we will 
consider $m_A$ to vary within the range of $10-90$ GeV, and we scan $\tan\beta$ between $20-100$. 
Lower vaues of $m_A$ are stronly constrained by $B_S\to \mu\mu$ processes. Note that limit on charged Higgs 
coming from $b\to s\gamma$ limits $\tan\beta >3$ \cite{Haller:2018nnx} for type-X 2HDM.

\subsection{CDF Result and Mass Spectrum}

\begin{figure}
\begin{center}
\includegraphics[width=7.5cm]{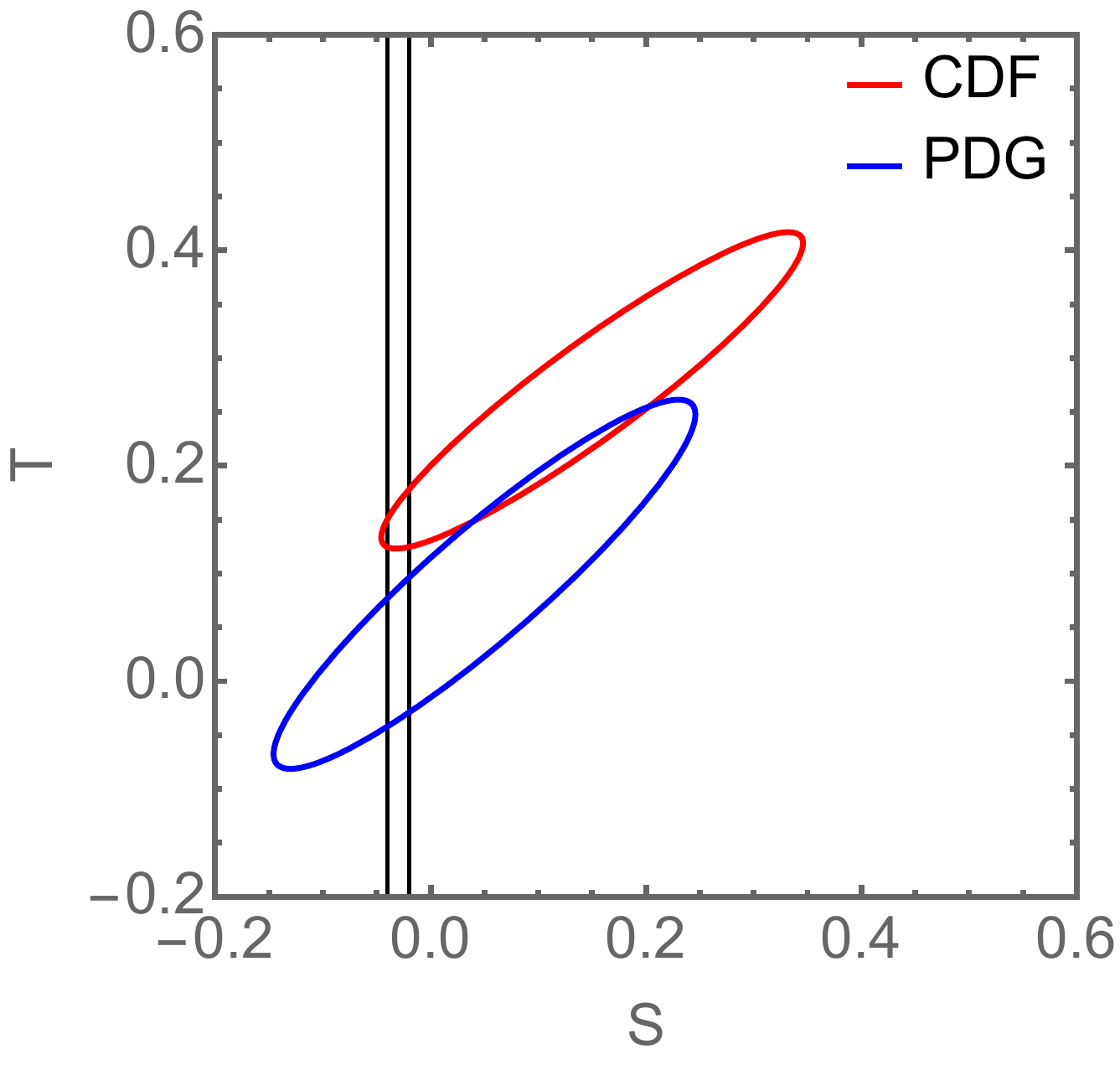}
\caption{ Preferred contours for the S and T oblique parameters with 2$\sigma$ C.L. 
The blue contour correspond to PDG data. The red contour correspond to data recently 
reported by the CDF II collaboration. The left and right vertical bars show the $S$ parameter when $m_H$ 
lies between $200$GeV and $400$GeV respectively. In type-X 2HDM model, $S$ is a negative correction. 
}\label{fig:ST}
\end{center}
\end{figure}
The quartic couplings in Eq.~\ref{eq:2hdm-pot} are constrained by the requirement of vacuum stability, 
perturbativity and unitarity. These constrains can be satisfied when~\cite{Wang:2014sda},
\bea
% m_{H} \simeq m_{H^\pm} &\leq& 250 \textrm{ GeV} \hspace{1cm} \textrm{(RS scenario)}\nonumber \\
m_{H} \simeq m_{H^\pm} &\leq& \sqrt{\la_{max}}\ v = \sqrt{4\pi}\ v.
\eea 
Here we assumed the wrong sign scenario i.e $\tan\beta \cos(\beta-\alpha) \sim 2$.

The constrains from precision measurement are usually defined in terms of S,T and U parameters. 
As we have discussed in the introduction, the effect of newly measured $m_W$ by CDF can be 
parametrized in terms of these parameters. Assuming $\Delta U=0$ the allowed values of S and T 
parameters are discussed in Ref.~\cite{Lu:2022bgw}, and the allowed range for S and T parameters are:
\begin{eqnarray}
	S &=& 0.15 \pm 0.08,~~~~~ T=0.27\pm 0.06.
\end{eqnarray}
\begin{figure}
\begin{center}
\includegraphics[width=7.5cm]{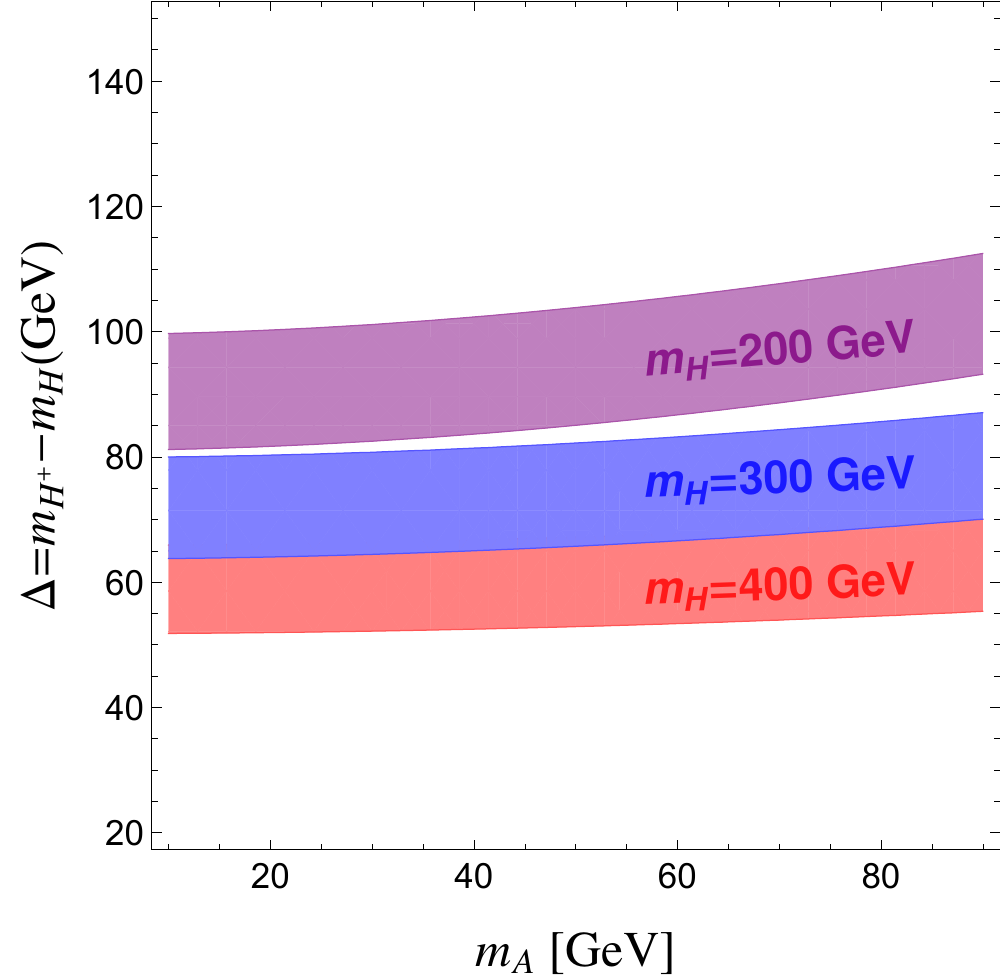}
\caption{Allowed range for $T$ parameter in the 2HDM with $m_H=200$ GeV (purple), $300$ GeV (blue), and $400$ GeV (Red). }
\label{fig:mdelta}
\end{center}
\end{figure}

In the 2HDM, the contributions to oblique parameters induced by the scalar bosons are given by \cite{Toussaint:1978zm}
\begin{widetext}
\begin{eqnarray}
S &=& -\frac{1}{4\pi} \left[ f_2\left( m_{H^\pm},m_{H^\pm}  \right)  -\sin^2\left(\beta -\alpha \right) f_2\left(m_H, m_A \right)  -\cos^2\left(\beta -\alpha \right) f_2\left(m_h, m_A \right)  \right],\nonumber\\
T &=& -\frac{1}{32\pi^2 \alpha_{EM} v^2} \left[ -f_1(m_A, m_{H^\pm}) +\sin^2\left(\beta -\alpha \right) \left( f_1\left(m_H, m_A \right) - f_1 \left( m_H, m_{H^\pm} \right) \right) \right.\nonumber\\
&& \left. \quad\quad\quad\quad\quad\quad  +\cos^2\left(\beta -\alpha \right) \left( f_1\left(m_h, m_A\right) - f_1\left( m_h, m_{H^\pm} \right) \right) \right],
\end{eqnarray}
where 
\begin{eqnarray}
f_1\left( m_1, m_2 \right) &=& \frac{m^2_1+m^2_2}{2} - \frac{m^2_1 m^2_2}{m^2_1 -m^2_2} \log\frac{m^2_1}{m^2_2}, \\
f_2\left( m_1, m_2 \right) &=&  -\frac{1}{3} \left( \frac{4}{3} - \frac{m^2_1 \log m^2_1 -m^2_2 \log m^2_2}{m^2_1 -m^2_2} - \frac{m^2_1 +m^2_2}{\left( m^2_1 - m^2_2 \right)^2 } f_1\left(m_1, m_2 \right) \right). 
\end{eqnarray}
\end{widetext}
%In the case of $m_A < m_Z \ll m_H < m_{H^\pm} $, the corrections are approximately given by
%\begin{eqnarray}
%	S &\sim& -\frac{5}{72\pi} \simeq -0.02,\\
%	T &\sim& \frac{m_H  (m_{H^\pm} - m_H) }{32\pi^2 \alpha_{EM} v^2} \simeq 0.01 \times \left( \frac{m_H}{200{\rm GeV}} \right) \left(  \frac{m_{H^\pm} -m_H}{10 {\rm GeV}} \right).
%\end{eqnarray}
The S parameter remains very small, and this strongly constrain the possible value of T parameter. 

We plotted the allowed space in S-T plane in Fig.~\ref{fig:ST} and the vertical bar 
shows the value of S parameter for heavy Higgs mass in the range of 200 to 400 GeV. As expected, 
only a small range of T parameter is allowed. We have satisfied the T parameter and found that 
the CP even heavy Higgs and the charged Higgs can not be degenerate. We plotted the mass differnce 
between $H$ and $H^\pm$ in Fig.~\ref{fig:mdelta}. The mass differnce has a mild dependence 
on the pseudoscalar mass.

We plotted the allowed space in the S-T plane in Fig.~\ref{fig:ST} and the vertical bar 
shows the value of the S parameter for heavy Higgs mass in the range of 200 to 400 GeV. As expected, 
only a small range in the T parameter is allowed. We have satisfied the T parameter and found that 
the CP even heavy Higgs and the charged Higgs can not be degenerate. We plotted the mass difference 
between $H$ and $H^\pm$ in Fig.~\ref{fig:mdelta}. The mass difference has a mild dependence 
on the pseudoscalar mass.

\subsection{Lepton Universality}

The lepton universality has been measured by $Z$ and $\tau$ decays at the level of $0.1\%$ \cite{ALEPH:2005ab,HFLAV:2016hnz}. 
In the type-X 2HDM, loop corrections provide huge contributions due to the enhanced leptonic couplings of extra Higgs bosons with large $\tan\beta$. 

Let us first consider the lepton universality from $Z$ boson decays. 
The $Z$ decay measurements can be converted into the leptonic branching ratios which reads
\begin{eqnarray}
\frac{ \Gamma\left(Z \to \mu^+\mu^- \right) }{ \Gamma\left(Z \to e^+e^- \right) } &=& 1.0009 \pm 0.0028,\nonumber\\
\frac{ \Gamma\left(Z \to \tau^+\tau^- \right) }{ \Gamma\left(Z \to e^+e^- \right) } &=& 1.0019 \pm 0.0032,
\end{eqnarray}
with a positive correlation of $0.63$.
For each lepton flavor, we can evaluate different quantities which can be parameterized as
\begin{eqnarray}
\delta_{\ell\ell} &\equiv& \frac{ \Gamma\left(Z \to \ell^+\ell^- \right) }{ \Gamma\left(Z \to e^+e^- \right) }-1.
\end{eqnarray}
A sizable contributions can generate from the tau Yukawa coupling of $m_\tau \tan\beta / v$. 

The other lepton universality has been evaluated by the pure and semi-hadronic decay processes which are given by
\begin{widetext}
\begin{eqnarray}
&&
\left( g_\tau \over g_\mu \right) = 1.0010 \pm 0.0015, \quad
\left( g_\tau \over g_e \right) = 1.0029 \pm 0.0015, \quad
\left( g_\mu \over g_e \right) = 1.0019 \pm 0.0014,  \nonumber\\
&&
\left( g_\tau \over g_\mu \right)_\pi = 0.9961 \pm 0.0027, \quad
\left( g_\tau \over g_\mu \right)_K= 0.9860 \pm 0.0070,
\end{eqnarray}
where the correlation matrix is
\begin{eqnarray}
\left(
\begin{array}{ccccc}
	1 & +0.53 & -0.49 & +0.24 & +0.11 \\
	+0.53  & 1     &  + 0.48 & +0.26    & +0.10 \\
	-0.49  & +0.48  & 1       &   +0.02 & -0.01 \\
	+0.24  & +0.26  & +0.02  &     1    &     +0.06 \\
	+0.11  & +0.10  & -0.01  &  +0.06  &   1 
\end{array} \right) .
\end{eqnarray}
\end{widetext}
In the large $\tan\beta$ limit, there are two important corrections.
One correction is originated from the heavy charged Higgs boson at the tree-level. 
The other one comes from the extra Higgs bosons through loop-level corrections.

So far in the literature the constraint on 2HDM-X coming from lepton universality
has been calculated~\cite{Abe:2015oca,Chun:2016hzs,Chun:2019oix} assuming that $H$ and $H^\pm$ degenerate.
However, the CDF result calls for a non-degenerate mass spectrum, and we have recalculated the lepton 
universality limits for such a scenario. We found that unless $H$ is relatively light ($m_H \sim 200$ GeV) 
the non-degenerate mass spectrum does not affect the lepton universality limits significantly. This is due 
to the fact that the lepton universality mostly depends on the mass gap between $m_A$ and $m_H$ when $H^\pm$  
is heavier.
Following the $\chi^2$-analysis in Ref.\cite{Chun:2016hzs,Chun:2018ibr}, we obtain the limits on the lepton universality of $\tau$ and $Z$ decays.

\begin{figure}
\begin{center}
\includegraphics[width=8.7cm]{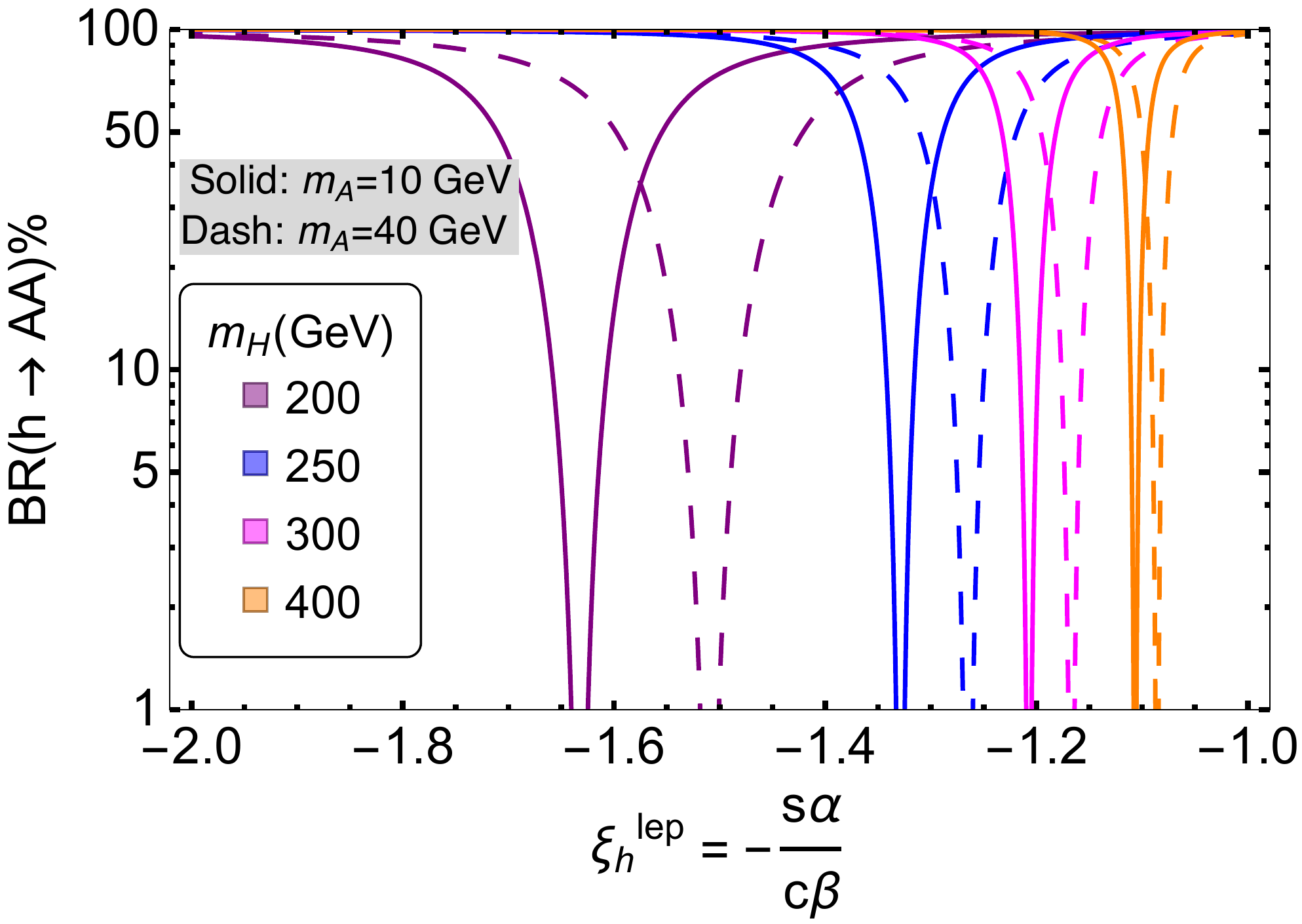}
\caption{Branching ratio of $h\to AA$ for different values of $m_A$ and $m_H$ as a function of the coupling modifier define in Tab.~\ref{Tab:YukawaFactors}.} 
\label{fig:BRhAA}
\end{center}
\end{figure}

\subsection{Constraints from Higss decay}
Since the pseudoscalar is lighter than $m_h/2$, the SM Higgs can decay to a pair of 
pseudoscalars. Since the pseudoscalar couples strongly to leptons, the relevant 
signal at the LHC is $p p \to h\to AA\to 4\tau/2\mu2\tau$. The both CMS and ATLAS have studied 
these channels \cite{ATLAS:2015unc, CMS:2017dmg, ATLAS:2018coo,CMS:2018qvj, CMS:2019spf,  CMS:2020ffa}. 
From these observations, the limit on $BR(h\to AA)$ is as low as $< 2\% $ for $m_A = 9$ GeV. 
For most of the $m_A$ range, the limit is $BR(h\to AA)<5\%$. We need to satisfy these constraints. 

The relevant coupling for the Higgs decay to $AA$ is given by \cite{Gunion:2002zf},
\bea
\lambda_{T}&=&\frac{1}{4}s_{2\beta}^2 (\la_1+\la_2)+(\la_3+\la_4+\la_5)(s_\beta^4+c_\beta^4)-2\la_5\nonumber\\
\lambda_{U}&=&\frac{1}{2}s_{2\beta} (s_\beta^2\la_1-c_\beta^2\la_2+c_{2\beta}(\la_3+\la_4+\la_5))\nonumber\\
\lambda_{hAA}&=&-v(\lambda_T s_{\beta-\alpha}-\lambda_Uc_{\beta-\alpha})
\eea
By rewriting $\sin\alpha = \xi_h^\ell \cos\beta$ and by inverting the quartic coupling to the 
relevant mass parameter we can calculate the branching ratio as a function of $m_H, m_A, \xi_h^\ell$. 
Figure.~\ref{fig:BRhAA} depictes the branching ratios. Evidently,  $BR(h\to AA)$  is close to $100\%$ for most of the parameter 
space, unless the parameters are fine-tuned to cancel. 
This cancellation happens for a particular value of coupling modifier depending on the mass of heavy Higgs. 
The supression in the branching ratio does not depend on the quartic coupling $\lambda_1$, another free parameter of the model. 
From the Figure, it is evident that the parameter space is very strongly constrained. Since the coupling 
modifier dictates the deviation of Higgs Yukawa coupling, it is constrained by the measurement 
of Higgs coupling strength. The current measurement of these coupling strengths by CMS and ATLAS collaboration 
allows $\xi_h^\ell$ to be within 0.7 to 1.15 at 2$\sigma$~\cite{CMS:2022kdi, ATLAS:2022yrq}. 
Thus, the combination of Higgs signal strength measurement with the measurement of Higgs branching
ratio to light pseudoscalar puts a lower limit on the mass spectrum of 2HDM-X, and the CP even Higgs has to 
heavier than 300 GeV.

\section{\label{sec:result}Numerical Results and Discussions}
\begin{figure}
\begin{center}
\includegraphics[width=7.5cm]{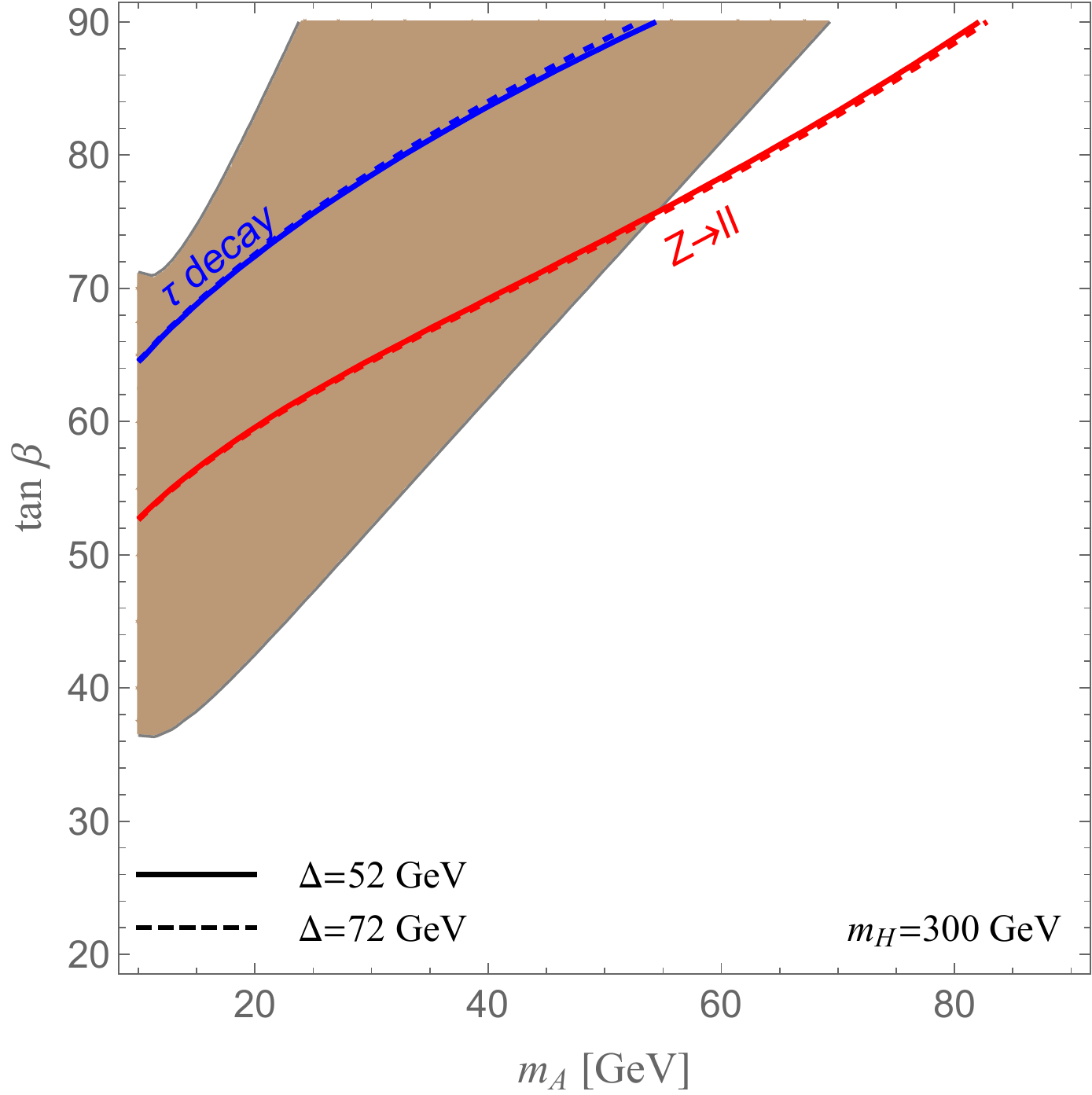}
\includegraphics[width=7.5cm]{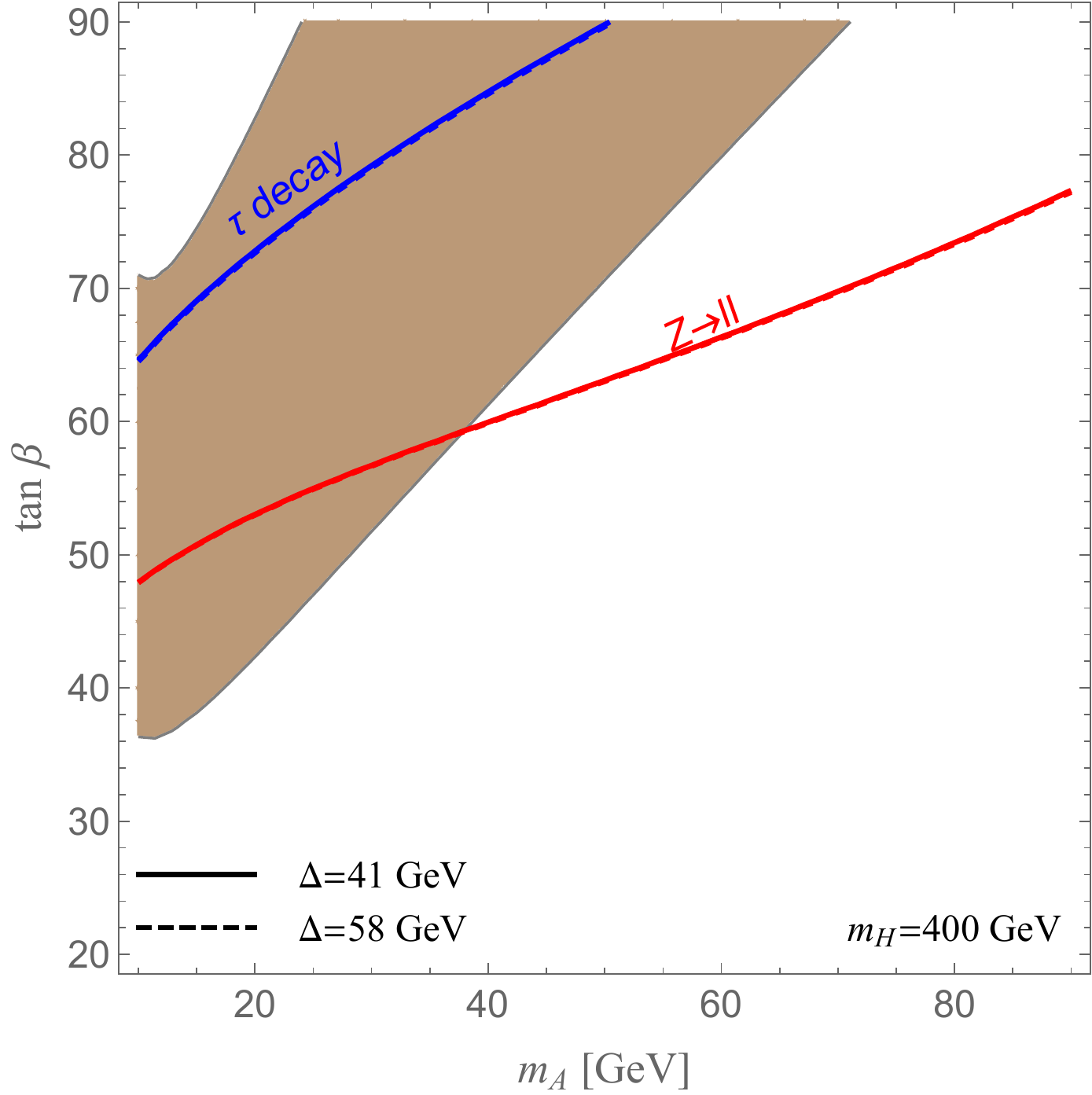}
\centering
\caption{Allowed parameter space in $(m_A, \tan\beta)$ plane describing all the constraints.
	The brown area is allowed by the muon $g-2$ at $2\sigma$ C.L. Below blue solid (dashed) 
	lines by the lepton universality test with $\tau$ decays are allowed. Below red solid 
	(dashed) lines by the lepton universality test with $Z$ decays are preffered. 
	We take $m_H=300~ {\rm (top)}$ and $m_H=400~ {\rm (bottom)}$GeV. Solid (dashed) lines correspond to the minimal (maximal) mass difference to resolve the CDF II $W$ boson mass anomaly.}
\label{fig:gm2}
\end{center}
\end{figure}
In the previous section we have discussed the relevant parameter space and the contraines coming from 
various theoretical as well as experimental observatios. We scaned the parmater space in the 
following range:
\begin{equation}
 m_A: 10-90~  \textrm{ GeV},~\tan\beta: 20-90,~m_H : 300~\&~ 400 \textrm{ GeV} 
\end{equation}
The allowed parameter and the constraints are shown in Fig.~\ref{fig:gm2} where the top(bottom) panel 
corresponds to $m_H = 300(400)$ GeV. To explain $\gmt$, we neet light pseudoscalar and large $\tan\beta$
and the parameter space is strongly constrained by the measurement of lepton universality. However, small parameter space
with $m_A$ between 10 to 50 GeV is still allowed when $\tan\beta$ is in the range $40-60$ GeV. 

Since the BSM scalars are hadrophobic, the present limit on the scalars of type-X 2HDM is rather weak. 
The existing LHC limits originate from the fermionic interactions of the BSM scalars, i.e. the existing 
search strategy looks for BSM scalars produced via gluon fusion and decay to $b\bar b$ or $\tau\tau$. 
Since the quark coupling to the new scalars is proportional to $\cot\beta$, there is no limit on BSM 
scalars if $\tan\beta$ is greater than 10. The parameter space of this model is well restricted, and 
the non-fermionic decay mode should be studied to probe the allowed parameter space.

\section{\label{sec:conclusion}Conclusion}
In this work, we explored type-X 2HDM to constrain the $(m_A, \tan\beta)$ parameter region with the recent measurement obtained by CDF II collaboration.
To resolve the W boson mass discrepancy, the mass difference between $m_H$ and $m_{H^\pm}$ are the order of $\mathcal{O}(10)$GeV due to enhancement in the 
oblique correction, $T \sim 0.1-0.2$. We calculated constraints coming from the lepton universality test and measurement of Higgs signal strength at the LHC. We found that the CDF measurement does not constrain the 2HDM-X parameter space, which can explain the muon $(g-2)$ anomaly. Interestingly, the Higgs signal strength 
measurement, specifically the measurement of $BR(h\to AA)$, put a strong lower limit on the mass of the heavy Higgs boson, which should be heavier than $200$GeV.% \section{Appendix}

\section*{Acknowledegments}
The work is supported in part by KIAS Individual Grant, Grant No. PG074202 (JK) at Korea Institute for Advanced Study.
I would like to thank Tanmoy Mondal for valuable comments and discussion on the related topics.

\bibliographystyle{utphys}
\bibliography{2hdm}

\end{document}